%% file: paper.tex
\documentclass[runningheads]{llncs}
\usepackage[T1]{fontenc}
\usepackage{graphicx}
\usepackage[dvipsnames]{xcolor}
\usepackage{pgfplots}
\pgfplotsset{compat=1.18}
\usepackage{amsmath,amsfonts,amssymb}
\numberwithin{table}{section}
\usepackage{stmaryrd}
\usepackage{import}
\usepackage[linesnumbered,ruled,vlined]{algorithm2e} 
\DontPrintSemicolon%

\usepackage[hidelinks,
    pdftitle={Revisiting Incremental Linearization for Nonlinear Integer Arithmetic},
    pdfauthor={Marek Danco, Karel Chvalovsky, Mikolas Janota}]{hyperref}
\usepackage[nameinlink,capitalize,noabbrev]{cleveref} 
\crefname{algocf}{Algorithm}{Algorithms}
\Crefname{algocf}{Algorithm}{Algorithms}

\newcommand{\vars}{\textsf{vars}}
\newcommand{\terms}{\textsf{terms}}

\newcommand{\dom}{\textsf{dom}}

\newcommand{\TRUE}{\textsf{T}}
\newcommand{\FALSE}{\textsf{F}}
\newcommand{\xs}{\bar{x}}

\newcommand{\solver}{\textsf{qfn2l}\xspace}

\newcommand{\model}[1]{\mu(#1)}
\newcommand{\pure}[1]{\llbracket #1 \rrbracket}

\title{Revisiting Incremental Linearization for Nonlinear Integer Arithmetic}
\author{Marek Dan\v{c}o\orcidID{0009-0008-3031-113X} \and
Karel Chvalovsk\'{y}\orcidID{0000-0002-0541-3889} \and
Mikol\'{a}\v{s} Janota\orcidID{0000-0003-3487-784X}}
\authorrunning{M. Dan\v{c}o et al.}
\institute{Czech Technical University in Prague}
\begin{document}
\maketitle

\begin{abstract}
	Incremental Linearization has previously been proposed for solving SMT problems over quantifier-free
	nonlinear integer arithmetic and has proven effective despite its conceptual simplicity.
	In this paper, we introduce a revised axiom set that improves convergence on polynomial constraints
	built from higher-degree monomials, such as powers and mixed products,
	a class of problems on which prior axiomatizations struggled. We present a standalone implementation
	built on top of Z3 for linear integer arithmetic and evaluate it on the NIA benchmark set from SMT-LIB.
	Our results show that the approach is competitive with state-of-the-art solvers overall and
	substantially outperforms them on benchmarks dominated by such polynomial constraints.
\end{abstract}

\input{intro}

\input{background}

\input{algorithm}
\input{axioms}
\input{experiments}
\input{comparison}
\input{conclusion}

\begin{credits}
\subsubsection{\ackname}
The research was supported by the European Union under the project
ROBOPROX (reg.\ no.\ CZ.02.01.01/00/22\_008/0004590) and by the Czech
Science Foundation grant no.~24-12759S.
This article is part of the RICAIP project that has received funding from the
European Union's Horizon 2020 research and innovation programme under grant
agreement No~857306.
\end{credits}

\bibliographystyle{splncs04}
\bibliography{refs}

\end{document}

%% file: intro.tex
\section{Introduction}

Quantifier-free nonlinear integer arithmetic (QF\_NIA) plays a central role
in automated reasoning, with applications ranging from program verification to
cryptographic protocol analysis. The satisfiability problem for QF\_NIA is
undecidable~\cite{matiyasevich1993hilbert}, which places fundamental limits on
any complete procedure and motivates the development of incomplete but
practically effective techniques.

The field of Satisfiability Modulo Theories (SMT) has seen remarkable pro\-gress,
with powerful and effective solvers now available for
linear integer arithmetic (LIA), linear real arithmetic (LRA), bit-vectors,
arrays, and their combinations.
Extending these advances to nonlinear arithmetic is considerably harder.
Moving from QF\_LIA to QF\_NIA introduces a fundamental barrier: not only is
the theory undecidable, but even its rational relaxation, QF\_NRA, requires
doubly exponential procedures such as Cylindrical Algebraic Decomposition
(CAD)~\cite{collins75} in the worst case.

A conceptually simple yet effective approach for dealing with nonlinear
arithmetic is \emph{incremental linearization}~\cite{cimatti2017,cimatti-tcl18,cimatti2018sat}.
The key idea is to replace each nonlinear operation with a fresh
\emph{uninterpreted function symbol}: a product $x y$ becomes
$f_\times(x, y)$, and integer division and modulo with non-constant divisors
are treated analogously.
This yields an abstraction in linear integer arithmetic with uninterpreted
functions (UFLIA).
When the UFLIA solver finds a model, it is checked
against the nonlinear semantics; if the check fails, linear \emph{axioms}
refining the uninterpreted symbols are added on demand, and the process is
repeated.

In this paper, we depart from the uninterpreted-function formulation and
instead abstract each nonlinear subterm by a plain fresh constant (a
\emph{pure}), working entirely within QF\_LIA\@.
The two abstractions differ in how they treat \emph{congruence}: a UFLIA
solver enforces it automatically, whereas with pures it must be enforced by
explicit axioms, added lazily when a violation is observed.
Our experiments indicate that such violations are rare (\Cref{sec:experiments}).

The performance of incremental linearization depends critically on the quality
of the added linear axioms: tight axioms reduce the number of iterations;
cheap axioms keep each iteration fast.
Prior work for QF\_NIA relied on a fixed collection of sign, zero, neutrality,
proportionality, and tangent-plane axioms for products~\cite{cimatti2018sat}.
While generally applicable, these axioms leave substantial room for improvement
on formulae dominated by \emph{polynomial} constraints, by which we mean
constraints built from higher-degree monomials, i.e., powers
$x^k$ and mixed products $x^k y^l$.
On such formulae, prior axiomatizations can struggle to converge within
practical time limits.

We revisit incremental linearization for QF\_NIA with a focus
on strengthening the axiom set for polynomial terms.
Our main contribution is a family of \emph{secant-based} linear bounds for
monomials $x^k$ and mixed products $x^k y^l$, derived from the observation
that on any unit integer interval $[v, v{+}1]$ the function $t \mapsto t^k$
admits tight piecewise-linear over- and under-approximations.
When instantiated at the current model values, these axioms provide
significantly tighter linearizations than prior general-purpose axioms,
enabling the solver to converge on hard polynomial benchmarks.

\begin{example}\label{ex:intro}
	Consider the following ``sum of three cubes'' benchmark\footnote{Benchmark
		\texttt{STC\_0079.smt2} from the \texttt{20220315-MathProblems} family in
		the SMT-LIB QF\_NIA suite.}:
	\[
		x^3 + y^3 + z^3 = 79.
	\]
	Such instances are a well-known challenge for SMT solvers, which must find
	a solution without any specialized knowledge of the number-theoretic
	structure. cvc5, MathSAT, Yices~2, and Z3 all time out on this instance
	within a 3-minute limit, while our solver finds the model
	$x = -19,\; y = 35,\; z = -33$ in around 20 seconds.
\end{example}

We implement the approach as a standalone solver,
\solver\footnote{\url{https://github.com/MarekDanco/qfn2l}}, built on
top of Z3 as the QF\_LIA backend.
An experimental evaluation on the full QF\_NIA benchmark suite from SMT-LIB
shows that \solver is competitive with state-of-the-art solvers overall
and substantially outperforms them on benchmark families dominated by
such polynomial constraints, most notably instances involving sums of cubes.

\paragraph{Contributions.}
Compared to prior work on incremental linearization for QF\_NIA,
we make the following contributions:
\begin{itemize}
	\item A new family of secant-based linear axioms for monomials $x^k$ and
	      mixed products $x^k y^l$, with a proof of soundness (\Cref{sec:axioms}).
	\item A complete description of our incremental linearization algorithm for
	      QF\_NIA, including purification and axiom selection (\Cref{sec:algorithm}).
	\item An experimental evaluation on the full QF\_NIA benchmark suite from
	      SMT-LIB, comparing against Z3, cvc5, Yices~2, and MathSAT
	      (\Cref{sec:experiments}).
\end{itemize}

%% file: background.tex
\section{Background}\label{sec:background}

In this paper, we assume the standard many-sorted first-order logic
with equality. We focus on satisfiability modulo theories (SMT) for
formulas in Quantifier-Free Nonlinear Integer Arithmetic (QF\_NIA) and
Quantifier-Free Linear Integer Arithmetic (QF\_LIA), optionally extended
with uninterpreted functions (QF\_UFLIA). Following the SMT-LIB~\cite{smtlib}
standard, the signature of QF\_NIA comprises integer constants (numerals),
functions $\{+, -, \cdot, \mathit{div}, \mathit{mod}, \mathit{abs}\}$, and
predicates $\{<, \leq, \approx, \geq, >\}$.
In QF\_LIA, multiplication is restricted to multiplication by integer
constants, and $\mathit{div}$ and $\mathit{mod}$ are restricted to nonzero
constant divisors; the remaining operators are unrestricted. In
QF\_UFLIA, arbitrary uninterpreted function symbols are also allowed.

Let $\vars$ be an infinite set of variables. We use
lowercase letters (e.g., $x, y, z$) to denote individual
variables and $\xs$ to denote a vector of variables. The set
$\terms$ of all terms, as well as atoms, literals, and
(sub)formulas are defined in the standard way. Formulas are
denoted $\varphi$.

A model (or interpretation) of a QF\_NIA (or QF\_LIA) formula $\varphi$, which shows that $\varphi$ is satisfiable,
has integers and \{\TRUE,\FALSE\} as its domains, and all the function and predicate symbols are
interpreted in the standard way. In QF\_UFLIA, a model also has to interpret the uninterpreted function symbols.

A substitution, denoted
$\mu\colon\vars\to\terms$, is a function that
assigns terms to variables. Here, we only consider its
(finite) nonidentity part
$\mu=\{x_1\mapsto t_1,\dots,x_n\mapsto t_n\}$, where
$x_i\neq t_i$, and $\dom(\mu)=\{x_1,\dots,x_n\}$. An
application of such a substitution $\mu$ to a formula
$\varphi$, denoted $\varphi[\mu]$, consists in replacing
all free variables $x_i$ by the corresponding term $t_i$
simultaneously.
By abuse of notation, $\mu$ is an assignment if all the terms $t_i$ are integer or Boolean constants with the standard interpretation.

A \emph{monomial} in variables $\xs=x_1,x_2,\dots,x_n$ is a product
$x_1^{k_1}x_2^{k_2}\cdots x_n^{k_n}$, where $k_i$ are nonnegative integers.
Monomials that involve a product of variables are the nonlinear terms axiomatized in \Cref{sec:axioms}.

%% file: algorithm.tex
\section{Algorithm}\label{sec:algorithm}

Our solver is an instance of counterexample-guided abstraction refinement
(CEGAR)~\cite{clarke2000cegar}: the input formula is abstracted into QF\_LIA,
and the abstraction is iteratively strengthened until it either becomes
unsatisfiable or yields a model correct under the nonlinear semantics.
Three issues arise: the abstraction forgets both the meaning of the
nonlinear operations and \emph{congruence} between their occurrences, so a
model of the abstraction may be spurious; the solver must decide which
abstracted terms to blame, which we do via an \emph{implicant} of the
abstract formula (a subset of its literals sufficient for the model to
satisfy it); and the offending terms must be refined by linear axioms
(\Cref{sec:axioms}) that exclude the spurious model.
Additionally, before returning to the LIA solver, we attempt a cheap
\emph{model repair} that tries to patch the spurious model directly.

\emph{Normalization and purification.}
Given a QF\_NIA formula $\varphi$, nonlinear monomials are first
\emph{normalized}: by introducing fresh variables with defining equalities,
every monomial is rewritten into the form $c \cdot x^k y^l$, where $c$ is an
integer constant and $x, y$ are variables.
For instance, $x y z$ becomes $d z$ for a fresh variable $d$ constrained by
$d \approx x y$; nonlinear factors other than variables, such as integer
divisions, are likewise replaced by fresh defined variables.
The second step is \emph{purification}: every nonlinear subterm $t$, i.e., a
monomial $x^k y^l$ with $k + l \geq 2$, or an integer division/modulo with a
non-constant divisor, is replaced by a fresh integer constant $\pure{t}$,
called a \emph{pure}.
Each monomial is replaced as a whole: $x y^2$ yields the single pure
$\pure{x y^2}$ and no pures for its sub-terms $x y$ or $y^2$.
Equal subterms share the same pure.
The result $\hat\varphi$ is a QF\_LIA formula over the original variables and
the new pure constants.

\emph{Main loop.}
\Cref{alg:solve} shows the top-level procedure.
We maintain a set $\mathcal{A}$ of linear \emph{axioms} accumulated across iterations.
Each iteration calls a QF\_LIA solver on $\hat\varphi \land \bigwedge\mathcal{A}$.
If the LIA problem is unsatisfiable, so is the original QF\_NIA formula.
Otherwise we obtain a model $\mu$ and call \textsc{Check-Nia} (\Cref{alg:checknia}),
which returns a set $\mathcal{N}$ of new axioms.
If $\mathcal{N} = \emptyset$, the current model $\mu$ is NIA-valid and we return~$\mathit{sat}$.
Otherwise we expand $\mathcal{A}$ with $\mathcal{N}$ and repeat.

\begin{algorithm}[t]
	\caption{\textsc{Qf-Nia-Solve}($\varphi$)}\label{alg:solve}
	\DontPrintSemicolon
	$\hat\varphi,\, P \leftarrow \textsc{Purify}(\varphi)$\;
	$\mathcal{A} \leftarrow \emptyset$\;
	\While{$\mathit{true}$}{
		$\mathit{res} \leftarrow \textsc{Check-Lia}(\hat\varphi \land \bigwedge\mathcal{A})$\;
		\lIf{$\mathit{res} = \mathit{unsat}$}{\Return $\mathit{unsat}$}
		$\mu \leftarrow \textsc{Model}(\mathit{res})$\;
		$\mathcal{N} \leftarrow \textsc{Check-Nia}(\varphi,\hat\varphi,P,\mu,\mathcal{A})$\;
		\lIf{$\mathcal{N} = \emptyset$}{\Return $\mathit{sat},\, \mu$}
		$\mathcal{A} \leftarrow \mathcal{A} \cup \mathcal{N}$\;
	}
\end{algorithm}

\begin{algorithm}[t]
	\caption{\textsc{Check-Nia}($\varphi, \hat\varphi, P, \mu, \mathcal{A}$)}\label{alg:checknia}
	\DontPrintSemicolon
	\lIf{$\mu \models_{\mathrm{NIA}} \varphi$}{\Return $\emptyset$}
	$I \leftarrow \textsc{Implicant}(\hat\varphi, \mu)$\;
	$R \leftarrow \{\,\pure{t} \in I \mid \mu(\pure{t}) \neq \mu(t)
		\wedge \pure{t}\text{ occurs in a falsified literal}\,\}$\;
	$\mathit{res}, \mathcal{N} \leftarrow \textsc{Model-Fix}(\mu, R)$\;
	\lIf{$\mathit{res}$}{\Return $\emptyset$}
	$\mathcal{N} \leftarrow \mathcal{N} \cup {}$ congruence axioms for $R$\;
	\ForEach{$p \in R$}{
		$\mathcal{N} \leftarrow \mathcal{N} \cup \textsc{Axioms}(p, \mu)$\;
	}
	\Return $\mathcal{N}$\;
\end{algorithm}

\emph{NIA check and implicant-based targeting.}
\textsc{Check-Nia} first checks whether the values that $\mu$ assigns to the
variables already satisfy $\varphi$, regardless of the values of the pures;
if so, we return $\emptyset$ immediately.
Otherwise we extract an \emph{implicant} $I$ of $\hat\varphi$ under $\mu$ by
descending through the formula structure, keeping all conjuncts of a
conjunction and one $\mu$-satisfied disjunct of a disjunction.
We then evaluate each literal of $I$ with each pure $\pure{t}$ replaced by
the value of $t$ under $\mu$, and collect into $R$ the pures with
$\mu(\pure{t}) \neq \mu(t)$ that occur in a falsified literal.

\emph{Model repair.}
When $R \neq \emptyset$ we first attempt \textsc{Model-Fix}: a heuristic that pins
non-relevant variables and runs a small number of sub-iterations on a
restricted LIA problem, generating and accumulating axioms along the way.
If a NIA-valid model is found, \textsc{Check-Nia} returns $\emptyset$ immediately.
If not, the axioms generated during model repair are retained and we fall
through to the standard axiom generation step.

\emph{Axiom generation.}
For each failing pure $\pure{t} \in R$, we add to $\mathcal{N}$ every
axiom in our axiom set for $t$ that is violated by $\mu$.
Axiom classes are described in \Cref{sec:axioms}; they cover sign and zero
conditions, secant-based linear bounds for monomials $x^k$, bounds for mixed
products $x^k y^l$, tangent-plane lemmas, and integer division and modulo
constraints.
Lazy \emph{congruence} axioms are collected into $\mathcal{N}$ before
per-pure axiom generation: for a pair of pures whose arguments agree, the
pures must be equal, e.g.\ $x_1 \approx x_2 \wedge y_1 \approx y_2
	\rightarrow \pure{x_1 y_1} \approx \pure{x_2 y_2}$; for equal powers
$\pure{x^k}$, $\pure{y^k}$ this is strengthened to sign-aware monotonicity
axioms.
For tangent-plane lemmas we optionally employ the \emph{frontier} strategy
of~\cite{cimatti2017tacas}: a per-pure bounding box tracks all model values
seen so far, and whenever a new point extends the box, extra tangent planes
are instantiated at the two mixed corners to ensure both upper and lower
bounds exist in every region.
A detailed comparison with the incremental linearization algorithm
of~\cite{cimatti2018sat} is given in \Cref{sec:comparison}.

%% file: axioms.tex
\section{Incremental Linearization of Nonlinear Terms}\label{sec:axioms}

\subsection{Terms of The Form $x^k$}

\input{tables/exp_ax}

Table~\ref{tab:pow-axioms} lists the axioms generated for $\pure{x^k}$ when $\mu(x) = v$ but
$\mu(\pure{x^k}) \neq v^k$.
The \textit{Eq-zero} and \textit{Eq} axioms enforce the correct value directly: the pure equals $v^k$
if and only if $x = v$ (or $x = -v$ for even $k$).
The \textit{Gap} axiom states that $\pure{x^k}$ cannot lie in the open gap between consecutive achievable values of $x^k$.
The \textit{Mod} axiom adds a modular congruence: $x \equiv v \pmod{m}$ implies
$x^k \equiv v^k \pmod{m}$; for $m = 2$ the converse holds as well.
By default we instantiate it only for $m = 2$.

\begin{figure}[ht]
	\centering
	\input{figures/secant}
	\caption{
		Blue lines show the functions $y=x$, $y=2x$, $y=3x$, and $y=4x$, which provide lower and upper linear bounds
		for $x^2$ on unit intervals between $0$ and $4$. For each $x$, the active bounds are the maximal lower bound
		and the minimal upper bound among all available bounds (shown as solid lines).}\label{fig:secant}
\end{figure}
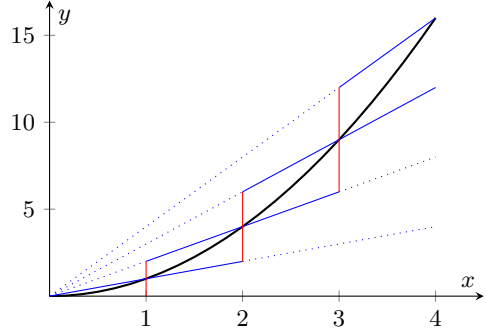

The \textit{Lin.\,LB} and \textit{Lin.\,UB} axioms provide secant-based linear bounds using the
line $v^{k-1}x$, which passes through the origin and the point $(v,\,v^k)$.
On each side of $v$ this line bounds $x^k$ from below or above (see \Cref{fig:secant}).
Tangent-plane lemmas, by contrast, only yield lower bounds for a convex monomial $x^k$;
the secant construction is therefore necessary to obtain two-sided constraints.

\begin{example}
Consider $x^2 \leq x \land x \geq 2$.
Purifying $x^2$ to a fresh integer $p$ gives the LIA abstraction $p \leq x \land x \geq 2$,
which is satisfiable.
Iteration~1 returns the model $x = 2$, $p = 1$; the NIA check fails since $2^2 = 4 \neq 1$,
and the \textit{Lin.\,LB} axiom at $v = 2$ is added:
\[
  x \geq 2 \;\implies\; p \geq 2x.
\]
Together with $p \leq x$ and $x \geq 2$ from the formula, the LIA solver derives
$2x \leq p \leq x$, hence $x \leq 0$, contradicting $x \geq 2$.
The solver returns \textit{unsat} after two iterations.
\end{example}

\subsection{Terms of The Form $x^k y^l$}

\input{tables/mul_ax_basic}
\input{tables/mul_ax_secant_conds}
\input{tables/mul_ax_secant_lb}
\input{tables/mul_ax_secant_ub}
\input{tables/mul_ax_tangent}

Table~\ref{tab:mul-axioms} lists the basic axioms for $\pure{x^k y^l}$, generated when
$\mu(x)=v$, $\mu(y)=w$, and $\mu(\pure{x^k y^l}) \neq v^k w^l$.
The \textit{Eq-zero} and \textit{Eq} axioms play the same role as in the $x^k$ case;
the \textit{Eq} axiom additionally exploits existing pures $\pure{x^k}$ or $\pure{y^l}$ when
available, replacing one factor by its pure rather than its model value.
The \textit{Mod} axiom states that $x \equiv v \pmod{m}$ and $y \equiv w \pmod{m}$ together
imply $x^k y^l \equiv v^k w^l \pmod{m}$; unlike the $x^k$ case, the converse does not hold in
general, not even for $m = 2$.

The \textit{Lin.\,LB} and \textit{Lin.\,UB} axioms
(Tables~\ref{tab:mul-axioms-lb}--\ref{tab:mul-axioms-ub}) linearize $x^k$ via the secant
conditions of Table~\ref{tab:secant-conds} while treating $w^l$ as a constant coefficient,
yielding the linear bound $v^{k-1}x w^l$.
The choice of $\downarrow$ or $\uparrow$ for each factor depends on the signs of $v^k$ and
$w^l$ so that the inequality direction is preserved under multiplication; the four cases are
proved below.

The \textit{Tangent} axioms (Table~\ref{tab:mul-tangent-axioms}) apply only to the bilinear
case $k = l = 1$.
The tangent plane of $xy$ at $(v, w)$ is $wx + vy - vw$, and the sign of
$xy - (wx + vy - vw) = (x-v)(y-w)$ determines whether $\pure{xy}$ lies above or below this
plane.
These axioms coincide with the McCormick envelope~\cite{mccormick1976} evaluated at the
model point; the secant-based bounds above extend the same idea to higher-degree monomials
$x^k y^l$.

By inspection of Table~\ref{tab:secant-conds}, the two conditions satisfy the inequality chains
\[
	\mathit{C}(x^k,v^k,\downarrow) \implies v^k \leq v^{k-1}x \leq x^k,
	\qquad
	\mathit{C}(x^k,v^k,\uparrow) \implies x^k \leq v^{k-1}x \leq v^k,
\]
so each condition simultaneously makes both $v^k$ and $v^{k-1}x$ lower (resp.\ upper) bounds on $x^k$,
with the linear bound $v^{k-1}x$ always the tighter of the two.

The validity of the lower-bound axioms (Table~\ref{tab:mul-axioms-lb}) follows from a two-step chain.
Under the conditions of any given row,
\[
	x^k y^l \;\geq\; x^k w^l \;\geq\; v^{k-1}x w^l.
\]

\emph{Step~1} (right inequality).
$\mathit{C}(x^k,v^k,{\cdot})$ bounds $x^k$ relative to $v^{k-1}x$.
Multiplying by $w^l$ preserves the direction when $w^l>0$ and reverses it when $w^l<0$.
The table compensates: it selects $\downarrow$ (i.e.\ $x^k \geq v^{k-1}x$) when $w^l>0$
and $\uparrow$ (i.e.\ $x^k \leq v^{k-1}x$) when $w^l<0$,
so that $x^k w^l \geq v^{k-1}x w^l$ in all four sign combinations.

\emph{Step~2} (left inequality).
$\mathit{C}(y^l,w^l,{\cdot})$ bounds $y^l$ relative to $w^l$.
Multiplying by $x^k$ preserves the direction when $x^k>0$ and reverses it when $x^k<0$.
Within the region defined by $\mathit{C}(x^k,v^k,{\cdot})$, the sign of $x^k$ matches the sign of $v^k$:
for even $k$ both are non-negative, and for odd $k$ the conditions in Table~\ref{tab:secant-conds}
confine $x$ and $v$ to the same side of zero.
The table selects $\downarrow$ (i.e.\ $y^l \geq w^l$) when $v^k>0$
and $\uparrow$ (i.e.\ $y^l \leq w^l$) when $v^k<0$,
so that $x^k y^l \geq x^k w^l$.

The upper-bound axioms (Table~\ref{tab:mul-axioms-ub}) follow from the symmetric chain
$x^k y^l \leq x^k w^l \leq v^{k-1}x w^l$,
with $\downarrow$ and $\uparrow$ swapped throughout. Note that similar axioms can be
constructed also by linearizing $y$ while fixing constant $x$.

\subsection{Integer Division and Modulo}

The pure $\pure{x \,\mathit{div}\, y}$ (and symmetrically $\pure{x \,\mathit{mod}\, y}$) is
axiomatized by two families of axioms, generated when $\mu(x) = v$, $\mu(y) = w$, and the
pure's model value disagrees with the correct result.

\emph{Fix-divisor.}
Pinning $y$ to its model value $w$ makes the expression linear in $x$:
\[
  y = w \;\implies\; \pure{x \,\mathit{div}\, y} = x \,\mathit{div}\, w,
  \qquad
  y = w \;\implies\; \pure{x \,\mathit{mod}\, y} = x \,\mathit{mod}\, w.
\]
The right-hand sides are QF\_LIA expressions since $w$ is a numeral, so the LIA solver can
reason about them directly.

\emph{Large-divisor.}
When $|y|$ exceeds the magnitude of the dividend $x$, the result is determined.
For $v \geq 0$:
\[
  x = v \land |y| > v \;\implies\; \pure{x \,\mathit{div}\, y} = 0,
  \qquad
  x = v \land |y| > v \;\implies\; \pure{x \,\mathit{mod}\, y} = v.
\]
For $v < 0$ the conclusions reflect SMT-LIB semantics, where $\mathit{mod}$ is always
non-negative:
\begin{align*}
  x = v \land |y| \geq |v| &\;\implies\; \pure{x \,\mathit{div}\, y} = \begin{cases} -1 & y > 0 \\ 1 & y < 0, \end{cases} \\
  x = v \land |y| > |v| &\;\implies\; \pure{x \,\mathit{mod}\, y} = |y| - |v|.
\end{align*}
These axioms rule out spurious model values when the LIA solver assigns a nonzero quotient
or an incorrect remainder for a small dividend.

%% file: tables/exp_ax.tex
\begin{table}[ht]
	\centering
	\setlength{\tabcolsep}{1.2em}
	\renewcommand{\arraystretch}{1.3}
	\begin{tabular}{@{}lll@{}}
		\hline
		\textbf{Name}     & \textbf{Condition}     & \textbf{Axiom}                                              \\
		\hline
		\textit{Sign}     & $k$ even               & $\pure{x^k} \geq 0$                                         \\
		                  & $k$ odd                & $x > 0 \iff \pure{x^k} > 0$                                 \\
		\hline
		\textit{Eq-zero}  & $v = 0$                & $(\pure{x^k} = 0) \iff (x = 0)$                             \\
		\hline
		\textit{Eq}       & $v \neq 0$, \,$k$ odd  & $(\pure{x^k} = v^k) \iff (x = v)$                           \\
		                  & $v \neq 0$, \,$k$ even & $(\pure{x^k} = v^k) \iff (x = v \lor x = -v)$               \\
		\hline
		\textit{Gap}      & $v \neq 0$, \,$k$ odd  & $\pure{x^k} \leq v^k \lor \pure{x^k} \geq (v+1)^k$          \\
		                  & $v \neq 0$, \,$k$ even & $\pure{x^k} \leq v^k \lor \pure{x^k} \geq (|v|+1)^k$        \\
		\hline
		\textit{Lin.\,LB} & $v \geq 0$             & $x \geq v \implies \pure{x^k} \geq v^{k-1}\,x$              \\
		                  & $v < 0$, \,$k$ even    & $x \leq v \implies \pure{x^k} \geq v^{k-1}\,x$              \\
		                  & $v < 0$, \,$k$ odd     & $v \leq x \leq 0 \implies \pure{x^k} \geq v^{k-1}\,x$       \\
		\hline
		\textit{Lin.\,UB} & $v \geq 0$             & $0 \leq x \leq v \implies \pure{x^k} \leq v^{k-1}\,x$       \\
		                  & $v < 0$, \,$k$ even    & $v \leq x \leq 0 \implies \pure{x^k} \leq v^{k-1}\,x$       \\
		                  & $v < 0$, \,$k$ odd     & $x \leq v \implies \pure{x^k} \leq v^{k-1}\,x$              \\
		\hline
		\textit{Mod}      & $v \neq 0$             & $x \equiv v \pmod m \implies \pure{x^k} \equiv v^k \pmod m$ \\
		\hline
	\end{tabular}
	\vspace{0.9em}
	\caption{%
		Axioms generated for $\pure{x^k}$. \textit{Sign} is model-independent and added on the first NIA check failure; the remaining axioms are generated when $\model{x}=v$ and $\model{\pure{x^k}}\neq v^k$.\label{tab:pow-axioms}}
\end{table}

%% file: figures/secant.tex
\begin{tikzpicture}
\begin{axis}[
    width=0.6\linewidth,
    height=0.45\linewidth,
    domain=0:4,
    xmax=4.5,
    ymax=17,
    xlabel={$x$},
    ylabel={$y$},
    axis lines=middle,
]

\addplot[black, thick] {x^2};

\addplot[blue, domain=3:4] {4*x};
\addplot[blue, dotted, domain=0:3] {4*x};

\addplot[blue, domain=2:4] {3*x};
\addplot[blue, dotted, domain=0:2] {3*x};

\addplot[blue, dotted, domain=3:4] {2*x};
\addplot[blue, domain=1:3] {2*x};
\addplot[blue, dotted, domain=0:1] {2*x};

\addplot[blue, domain=0:2] {x};
\addplot[blue, dotted, domain=2:4] {x};

\addplot[
    red,
] coordinates {(1,0) (1,2)};
\addplot[
    red,
] coordinates {(2,2) (2,6)};
\addplot[
    red,
] coordinates {(3,6) (3,12)};

\end{axis}
\end{tikzpicture}

%% file: tables/mul_ax_basic.tex
\begin{table}[ht]
	\centering
	\setlength{\tabcolsep}{1.2em}
	\renewcommand{\arraystretch}{1.3}
	\begin{tabular}{@{}lll@{}}
		\hline
		\textbf{Name}    & \textbf{Condition}               & \textbf{Axiom}                                        \\
		\hline
		\textit{Sign}    & $k,\,l$ both even                & $\pure{x^k y^l} \geq 0$                               \\
		                 & $k$ even,\;$l$ odd               & $(x \neq 0 \land y > 0) \iff \pure{x^k y^l} > 0$      \\
		                 & $k$ odd,\;$l$ even               & $(x > 0 \land y \neq 0) \iff \pure{x^k y^l} > 0$      \\
		                 & $k,\,l$ both odd                 & $(x{>}0 \land y{>}0) \lor (x{<}0 \land y{<}0)$        \\
		                 &                                  & $\iff \pure{x^k y^l} > 0$                             \\
		\hline
		\textit{Eq-zero} & $v = 0$                          & $x = 0 \implies \pure{x^k y^l} = 0$                   \\
		                 & $w = 0$                          & $y = 0 \implies \pure{x^k y^l} = 0$                   \\
		\hline
		\textit{Eq}      & $v \neq 0$,\ $\pure{y^l}$ exists & $x = v \implies \pure{x^k y^l} = v^k \pure{y^l}$      \\
		                 & $w \neq 0$,\ $\pure{x^k}$ exists & $y = w \implies \pure{x^k y^l} = w^l \pure{x^k}$      \\
		                 & otherwise                        & $x = v \land y = w \implies \pure{x^k y^l} = v^k w^l$ \\
		\hline
		\textit{Mod}     & $v \neq 0,\ w \neq 0$            & $x \equiv v \pmod{m} \land y \equiv w \pmod{m}$       \\
		                 &                                  & $\implies \pure{x^k y^l} \equiv v^k w^l \pmod{m}$     \\
		\hline
	\end{tabular}
	\vspace{0.9em}
	\caption{%
		Basic axioms for $\pure{x^k y^l}$. \textit{Sign} is model-independent and added on the first NIA check failure; the remaining axioms are generated when $\model{x}=v$, $\model{y}=w$ and $\model{\pure{x^k y^l}}\neq v^kw^l$.\label{tab:mul-axioms}}
\end{table}

%% file: tables/mul_ax_secant_conds.tex
\begin{table}[ht]
	\centering
	\setlength{\tabcolsep}{1.2em}
	\renewcommand{\arraystretch}{1.3}
	\begin{tabular}{@{}lll@{}}
		\hline
		\textbf{Condition}      & $\mathit{C}(x^k,v^k,\downarrow)$ & $\mathit{C}(x^k,v^k,\uparrow)$ \\
		\hline
		$v \geq 0$         & $x \geq v$                   & $0 \leq x \leq v$          \\
		$v < 0$,\ $k$ even & $x \leq v$                   & $v \leq x \leq 0$          \\
		$v < 0$,\ $k$ odd  & $v \leq x \leq 0$            & $x \leq v$                 \\
		\hline
	\end{tabular}
	\vspace{0.9em}
	\caption{%
		Conditions $\mathit{C}(x^k,v^k,\downarrow)$ and $\mathit{C}(x^k,v^k,\uparrow)$ under which
		$v^{k-1}x$ and $v^k$ are lower (resp.\ upper) bounds on $x^k$,
		given model value $\model{x}=v$.\label{tab:secant-conds}}
\end{table}

%% file: tables/mul_ax_secant_lb.tex
\begin{table}[ht]
	\centering
	\setlength{\tabcolsep}{1.2em}
	\renewcommand{\arraystretch}{1.3}
	\begin{tabular}{@{}lll@{}}
		\hline
		\textbf{Name}    & \textbf{Condition}        & \textbf{Axiom}                                                                                          \\
		\hline
		\textit{Lin. LB} & $v^k > 0, \,w^l > 0$ & $\mathit{C}(x^k, v^k,\downarrow) \land \mathit{C}(y^l, w^l,\downarrow) \implies \pure{x^k y^l} \geq v^{k-1}xw^l$ \\
		                 & $v^k > 0, \,w^l < 0$ & $\mathit{C}(x^k, v^k,\uparrow) \land \mathit{C}(y^l, w^l,\downarrow) \implies \pure{x^k y^l} \geq v^{k-1}xw^l$   \\
		                 & $v^k < 0, \,w^l > 0$ & $\mathit{C}(x^k, v^k,\downarrow) \land \mathit{C}(y^l, w^l,\uparrow) \implies \pure{x^k y^l} \geq v^{k-1}xw^l$   \\
		                 & $v^k < 0, \,w^l < 0$ & $\mathit{C}(x^k, v^k,\uparrow) \land \mathit{C}(y^l, w^l,\uparrow) \implies \pure{x^k y^l} \geq v^{k-1}xw^l$     \\
		\hline
	\end{tabular}
	\vspace{0.9em}
	\caption{%
		\textit{Linear Lower Bound} axioms generated for $\pure{x^k y^l}$ when $\model{x}=v$, $\model{y}=w$, $\model{\pure{x^k y^l}}\neq v^kw^l$ and
		$x$ is linearized while $y$ is kept constant.
		Conditions $\mathit{C}(\cdot,\cdot,\downarrow/\uparrow)$ are defined in Table~\ref{tab:secant-conds}.\label{tab:mul-axioms-lb}}
\end{table}

%% file: tables/mul_ax_secant_ub.tex
\begin{table}[ht]
	\centering
	\setlength{\tabcolsep}{1.2em}
	\renewcommand{\arraystretch}{1.3}
	\begin{tabular}{@{}lll@{}}
		\hline
		\textbf{Name}    & \textbf{Condition}        & \textbf{Axiom}                                                                                          \\
		\hline
		\textit{Lin. UB} & $v^k > 0, \,w^l > 0$ & $\mathit{C}(x^k, v^k,\uparrow) \land \mathit{C}(y^l, w^l,\uparrow) \implies \pure{x^k y^l} \leq v^{k-1}xw^l$     \\
		                 & $v^k > 0, \,w^l < 0$ & $\mathit{C}(x^k, v^k,\downarrow) \land \mathit{C}(y^l, w^l,\uparrow) \implies \pure{x^k y^l} \leq v^{k-1}xw^l$   \\
		                 & $v^k < 0, \,w^l > 0$ & $\mathit{C}(x^k, v^k,\uparrow) \land \mathit{C}(y^l, w^l,\downarrow) \implies \pure{x^k y^l} \leq v^{k-1}xw^l$   \\
		                 & $v^k < 0, \,w^l < 0$ & $\mathit{C}(x^k, v^k,\downarrow) \land \mathit{C}(y^l, w^l,\downarrow) \implies \pure{x^k y^l} \leq v^{k-1}xw^l$ \\
		\hline
	\end{tabular}
	\vspace{0.9em}
	\caption{%
		\textit{Linear Upper Bound} axioms generated for $\pure{x^k y^l}$ when $\model{x}=v$, $\model{y}=w$, $\model{\pure{x^k y^l}}\neq v^kw^l$ and
		$x$ is linearized while $y$ is kept constant.
		Conditions $\mathit{C}(\cdot,\cdot,\downarrow/\uparrow)$ are defined in Table~\ref{tab:secant-conds}.\label{tab:mul-axioms-ub}}
\end{table}

%% file: tables/mul_ax_tangent.tex
\begin{table}[ht]
	\centering
	\setlength{\tabcolsep}{1.2em}
	\renewcommand{\arraystretch}{1.3}
	\begin{tabular}{@{}ll@{}}
		\hline
		\textbf{Name}    & \textbf{Axiom}                                        \\
		\hline
		\textit{Tangent} & $x > v \land y < w \implies \pure{xy} < wx + vy - vw$ \\
		                 & $x < v \land y > w \implies \pure{xy} < wx + vy - vw$ \\
		                 & $x < v \land y < w \implies \pure{xy} > wx + vy - vw$ \\
		                 & $x > v \land y > w \implies \pure{xy} > wx + vy - vw$ \\
		\hline
	\end{tabular}
	\vspace{0.9em}
	\caption{%
		\textit{Tangent} axioms generated for $\pure{xy}$ when $\model{x}=v$, $\model{y}=w$ and $\model{\pure{xy}}\neq vw$.\label{tab:mul-tangent-axioms}}
\end{table}

%% file: experiments.tex
\section{Experiments}\label{sec:experiments}

\emph{Setup.} All experiments were run on a server with four AMD EPYC 7513 32-core processors at 2.6\,GHz
and 504\,GB of memory.
Each instance was allocated a wall-clock timeout of 180\,seconds and a memory limit of 40\,GB.
We evaluated our solver, \solver, against Z3~4.15.4~\cite{z3}, cvc5~1.3.2~\cite{cvc5}, MathSAT5~5.6.16~\cite{mathsat5},
and Yices~2.6.4~\cite{Dutertre:cav2014} on the QF\_NIA benchmark set from SMT-LIB~\cite{smtlib}.
\solver is implemented in C++ using the smt-switch abstraction layer~\cite{mann2021smt},
linked against the same Z3~4.15.4 build used as the standalone Z3 baseline,
so any difference in results is due to the solving strategy rather than the underlying LIA engine.

Table~\ref{tab:results} summarises the results.
We report three configurations of \solver: the base solver (\solver), the base solver with
the frontier strategy for tangent-plane lemmas (\solver+\textsc{Frontier}), and an ablation
with lazy congruence axioms enabled (\solver+\textsc{Congr.}); the first two run without
congruence.
All three configurations include the secant-based axioms of \Cref{sec:axioms};
the frontier strategy (\Cref{sec:algorithm}) is an orthogonal tangent-plane
instantiation heuristic adopted from~\cite{cimatti2017tacas}.

\input{tables/results}

\input{tables/families}

On the full benchmark set \solver is competitive with cvc5 and trails Z3, MathSAT, and
Yices. Z3 combines a large number of built-in tactics and heuristics; MathSAT is
closed source. \solver, by contrast, is a concise open-source implementation with a
single LIA backend and no problem-specific tuning.
Since MathSAT5 implements the incremental linearization approach of Cimatti et
al.~\cite{cimatti2018sat} (cf.\ \Cref{sec:comparison}), the comparison with
MathSAT5 also serves as an experimental comparison with that work.

Enabling the frontier strategy adds a small but consistent improvement at no cost.
Enabling congruence axioms (\solver+\textsc{Congr.}) reduces the overall count by around
1\,600 instances, confirming that the overhead of pairwise congruence checks outweighs
their benefit on the general benchmark set.

The gap on satisfiable instances is most visible in large families such as ITS, AProVE, and
SAT14 (Table~\ref{tab:families}), and is largely structural: \solver must witness satisfiability
by iterative axiom refinement over a LIA abstraction, and each candidate solution is validated
purely through linear arithmetic.
Competing solvers employ complementary techniques, such as bit-blasting to fixed-width
integers, that can more directly enumerate satisfying
assignments for large-variable instances without the overhead of abstraction-refinement.
These families are particularly amenable to bit-blasting: the nonlinear structure is shallow,
instances are large (routinely hundreds of pures), and satisfying assignments tend to involve
small values.

The \texttt{20220315-MathProblems} family consists of 1100 instances encoding number-theoretic
problems, many of which involve sums of cubes ($x^3 + y^3 + z^3 = n$).
\solver solves 585--587 of these (53\%). This gap directly reflects the contribution of the
secant-based axioms for $x^k$: the convergence on cubic constraints requires bounds that go
beyond tangent-plane lemmas.

%% file: tables/results.tex
\begin{table}[t]
	\centering
	\setlength{\tabcolsep}{0.8em}
	\renewcommand{\arraystretch}{1.2}
	\begin{tabular}{lrrr|rrr}
		\hline
		                          & \multicolumn{3}{c|}{\textbf{Total} (25\,444)} & \multicolumn{3}{c}{\textbf{MathProblems} (1\,100)}                                                                     \\
		\textbf{Solver}           & \textbf{Sat}                                  & \textbf{Unsat}                                     & \textbf{Solved} & \textbf{Sat} & \textbf{Unsat} & \textbf{Solved} \\
		\hline
		\solver+\textsc{Frontier} & 8\,377                                        & 5\,709                                             & 14\,086         & 579          & 7              & 586             \\
		\solver                   & 8\,331                                        & 5\,680                                             & 14\,011         & 578          & 7              & 585             \\
		\solver+\textsc{Congr.}   & 6\,896                                        & 5\,525                                             & 12\,421         & 580          & 7              & 587             \\
		\hline
		Z3~4.15.4                 & 12\,718                                       & 6\,487                                             & 19\,205         & 111          & 7              & 118             \\
		MathSAT5~5.6.16           & 11\,177                                       & 5\,200                                             & 16\,377         & 164          & 7              & 171             \\
		Yices~2.6.4               & 10\,672                                       & 5\,181                                             & 15\,853         & 112          & 7              & 119             \\
		cvc5~1.3.2                & 8\,550                                        & 4\,446                                             & 12\,996         & 150          & 7              & 157             \\
		\hline
	\end{tabular}
	\vspace{0.9em}
	\caption{Instances solved within 180\,s on the SMT-LIB QF\_NIA benchmark set.\label{tab:results}}
\end{table}

%% file: tables/families.tex
\begin{table}[t]
    \centering
    \resizebox{\linewidth}{!}{%
    \begin{tabular}{l | lll | llll}
        \hline
        \textbf{Family} & \textbf{\solver+\textsc{Fr.}} & \textbf{\solver} & \textbf{\solver+\textsc{Cg.}} & \textbf{Z3} & \textbf{MathSAT} & \textbf{Yices} & \textbf{cvc5} \\
        \hline
        ITS (17046) & 4203/3908 & 4197/3881 & 3016/3738 & 8154/4452 & 6777/3569 & 6446/3434 & 4957/2843 \\
        AProVE (2409) & 1418/579 & 1397/579 & 1262/561 & 1642/687 & 1608/557 & 1590/708 & 1405/601 \\
        SAT14 (1926) & 1568/61 & 1566/61 & 1536/62 & 1853/72 & 1722/66 & 1809/65 & 1573/72 \\
        CInteger (1818) & 387/580 & 373/576 & 294/575 & 704/655 & 652/435 & 503/448 & 262/368 \\
        ReachSafety-Loops (350) & 9/310 & 9/310 & 9/317 & 11/339 & 10/320 & 6/302 & 11/322 \\
        mcm (186) & \textbf{11/0} & \textbf{12/0} & \textbf{11/0} & 8/0 & 7/0 & 6/0 & 9/0 \\
        calypto (177) & \textbf{80/97} & \textbf{80/97} & \textbf{80/97} & 79/97 & 79/90 & 79/95 & 79/96 \\
        leipzig (167) & 96/1 & 94/1 & 81/1 & 131/1 & 127/2 & 100/1 & 78/2 \\
        LassoRanker (106) & 4/91 & 4/92 & 4/91 & 4/100 & 4/101 & 4/85 & 4/92 \\
        UltimateAutomizerSvcomp2023 (58) & \textbf{7/15} & \textbf{7/15} & \textbf{7/15} & 8/12 & 7/10 & 5/0 & 6/0 \\
        UltimateLassoRanker (32) & 6/26 & 6/26 & 6/26 & 6/26 & 6/26 & 6/26 & 6/26 \\
        sqrtmodinv-hoenicke (27) & 0/17 & 0/17 & 0/17 & 0/17 & 0/0 & 0/0 & 0/1 \\
        ConcurrencySafety-Main (24) & 5/8 & 5/9 & 5/9 & 2/14 & 7/9 & 0/3 & 4/8 \\
        elster (9) & 4/0 & 3/0 & 5/0 & 5/0 & 7/0 & 6/0 & 6/0 \\
        UltimateAutomizer (7) & 0/7 & 0/7 & 0/7 & 0/7 & 0/7 & 0/7 & 0/7 \\
        LCTES (2) & \textbf{0/2} & \textbf{0/2} & \textbf{0/2} & 0/1 & 0/1 & 0/0 & 0/1 \\
        \hline
    \end{tabular}%
    }
    \vspace{0.9em}
    \caption{Sat/unsat instances solved per benchmark family within 180\,s (excluding \texttt{20220315-MathProblems}, see Table~\ref{tab:results}).\label{tab:families}}
\end{table}

%% file: comparison.tex
\section{Comparison with Prior Work}\label{sec:comparison}

We now compare our algorithm with the incremental linearization approach of
Cimatti et al.~\cite{cimatti2018sat}, implemented in MathSAT5.
Both instantiate the same CEGAR skeleton; the differences lie in the abstract
domain, axiom generation, failing-term selection, and model repair.

\emph{Abstract domain.}
Cimatti et al.~\cite{cimatti2018sat} replace each product $x y$ with an uninterpreted
function symbol $f(x,y)$, yielding a QF\_UFLIA abstraction; congruence
($f(a,b) = f(a',b')$ whenever $a=a'$ and $b=b'$) is then automatic.\footnote{This also means the Ackermann encoding~\cite{ackermann54} can reduce any QF\_UFLIA formula to QF\_LIA, so UF over integers does not add expressive power.}
We use plain fresh constants instead, so the abstract domain is QF\_LIA, and
congruence must be added as explicit axioms when violated.

\emph{Axiom generation.}
Cimatti et al.~\cite{cimatti2018sat} generate lemmas in three sequential rounds (basic axioms,
proportionality, tangent-plane), stopping as soon as a round produces at
least one lemma.
We add all violated axioms in a single pass, regardless of type.

\emph{Failing-term selection.}
Cimatti et al.~\cite{cimatti2018sat} scan all constraints of $\varphi$ falsified by $\mu$ and
collect every product $x y$ in those constraints.
We use an implicant of $\hat\varphi$ under $\mu$ and restrict to pures in
literals that fail under NIA semantics, which is more targeted.

\emph{Model repair.}
Cimatti et al.~\cite{cimatti2018sat} generate axioms and immediately return to the main UFLIA solve.
We first attempt \textsc{Model-Fix}: a heuristic sub-loop that pins irrelevant variables,
accumulates axioms over several restricted LIA calls, and tries to find a
NIA-valid model without returning to the outer loop.
When successful this saves an outer LIA call; when not, the axioms gathered
during the attempt are still retained and the outer loop continues as normal.

\emph{Axiom set.}
The axioms in~\cite{cimatti2018sat} cover sign, zero, neutrality,
proportionality, and tangent-plane conditions for binary products.
We retain sign, zero, and tangent-plane axioms and add secant-based bounds for
monomials $x^k$ and mixed products $x^k y^l$, described in \Cref{sec:axioms}.

%% file: conclusion.tex
\section{Conclusion and Future Work}\label{sec:conclusion}
We presented a novel incremental linearization approach for quantifier-free
nonlinear integer arithmetic.
Since the problem is undecidable in general, no complete procedure exists;
we therefore focused on techniques that perform well on benchmarks arising
in practice.
The key distinguishing feature of our approach is the explicit treatment of
power terms with fixed exponents, such as $x^3$, as first-class abstractions.
This yields a measurable advantage on problems where such expressions occur,
as confirmed by our experimental evaluation against state-of-the-art solvers.

Several directions remain for future work.
We plan to add a bit-blasting~\cite{bitblast} preprocessing phase to quickly discharge
satisfiable instances whose solutions are of small magnitude.
We also intend to prune axioms that become redundant upon the addition of
stronger ones, and to develop new axiom schemas targeting specific classes
of problems.


%% file: refs.bib
@inproceedings{bitblast,
  author        = {Jia, Fuqi and Han, Rui and Huang, Pei and Liu, Minghao and Ma, Feifei and Zhang, Jian},
  title         = {Improving Bit-Blasting for Nonlinear Integer Constraints},
  year          = {2023},
  isbn          = {9798400702211},
  publisher     = {{ACM}},
  booktitle     = {International Symposium on Software Testing and Analysis, {ISSTA}},
  pages         = {14--25}
}

@inbook{cimatti2017,
  title         = {Satisfiability Modulo Transcendental Functions via Incremental Linearization},
  isbn          = {9783319630465},
  issn          = {1611-3349},
  booktitle     = {Automated Deduction – CADE 26},
  publisher     = {Springer},
  author        = {Cimatti,  Alessandro and Griggio,  Alberto and Irfan,  Ahmed and Roveri,  Marco and Sebastiani,  Roberto},
  year          = {2017},
  pages         = {95–113}
}

@inproceedings{z3,
  author        = {Leonardo Mendon{\c{c}}a de Moura and Nikolaj Bj{\o}rner},
  title         = {{Z3:} An Efficient {SMT} Solver},
  booktitle     = {Tools and Algorithms for the Construction and Analysis of Systems, {TACAS}},
  pages         = {337--340},
  year          = {2008},
  crossref      = {DBLP:conf/tacas/2008},
  opturl        = {https://doi.org/10.1007/978-3-540-78800-3\_24},
  bibsource     = {dblp computer science bibliography, https://dblp.org}
}

@inproceedings{cvc5,
  author        = {Haniel Barbosa and Clark W. Barrett and Martin Brain and Gereon Kremer and Hanna Lachnitt and others},
  opteditor     = {Dana Fisman and Grigore Rosu},
  title         = {cvc5: {A} Versatile and Industrial-Strength {SMT} Solver},
  booktitle     = {Tools and Algorithms for the Construction and Analysis of Systems, {TACAS}},
  series        = {LNCS},
  volume        = {13243},
  pages         = {415--442},
  publisher     = {Springer},
  year          = {2022},
  opturl        = {https://doi.org/10.1007/978-3-030-99524-9\_24},
  bibsource     = {dblp computer science bibliography, https://dblp.org}
}

@book{ackermann54,
  title         = {Solvable Cases of the Decision Problem},
  author        = {Ackermann, Wilhelm},
  publisher     = {North-Holland},
  year          = {1954},
  series        = {Studies in Logic and the Foundations of Mathematics}
}

@article{cimatti-tcl18,
  author        = {Alessandro Cimatti and Alberto Griggio and Ahmed Irfan and Marco Roveri and Roberto Sebastiani},
  title         = {Incremental Linearization for Satisfiability and Verification Modulo Nonlinear Arithmetic and Transcendental Functions},
  journal       = {{ACM} Trans. Comput. Log.},
  volume        = {19},
  number        = {3},
  pages         = {19:1--19:52},
  year          = {2018},
  bibsource     = {dblp computer science bibliography, https://dblp.org}
}

@inproceedings{collins75,
  author        = "Collins, George E.",
  title         = "Quantifier elimination for real closed fields by cylindrical algebraic decomposition",
  booktitle     = "Automata Theory and Formal Languages",
  year          = "1975",
  publisher     = "Springer",
  pages         = "134--183",
  isbn          = "978-3-540-37923-2"
}

@misc{smtlib,
  author        = {Clark Barrett and Pascal Fontaine and Cesare Tinelli},
  title         = {{The Satisfiability Modulo Theories Library (SMT-LIB)}},
  howpublished  = {{\tt www.SMT-LIB.org}},
  year          = 2016
}

@proceedings{DBLP:conf/tacas/2008,
  opteditor     = {C. R. Ramakrishnan and Jakob Rehof},
  title         = {Tools and Algorithms for the Construction and Analysis of Systems, 14th International Conference, {TACAS} 2008},
  optseries     = {LNCS},
  volume        = {4963},
  publisher     = {Springer},
  year          = {2008},
  opturl        = {https://doi.org/10.1007/978-3-540-78800-3},
  isbn          = {978-3-540-78799-0},
  bibsource     = {dblp computer science bibliography, https://dblp.org}
}

@book{matiyasevich1993hilbert,
  title         = {Hilbert's Tenth Problem},
  author        = {Matiyasevich, Y.V.},
  isbn          = {9780262132954},
  lccn          = {lc93028107},
  series        = {Foundations of computing},
  year          = {1993},
  publisher     = {MIT Press}
}

@inproceedings{cimatti2018sat,
  author        = "Cimatti, Alessandro and Griggio, Alberto and Irfan, Ahmed and Roveri, Marco and Sebastiani, Roberto",
  title         = "Experimenting on Solving Nonlinear Integer Arithmetic with Incremental Linearization",
  booktitle     = "Theory and Applications of Satisfiability Testing -- SAT 2018",
  year          = "2018",
  publisher     = "Springer",
  pages         = "383--398",
  isbn          = "978-3-319-94144-8"
}

@inproceedings{cimatti2017tacas,
  author        = "Cimatti, Alessandro and Griggio, Alberto and Irfan, Ahmed and Roveri, Marco and Sebastiani, Roberto",
  title         = "Invariant Checking of {NRA} Transition Systems via Incremental Reduction to {LRA} with {EUF}",
  booktitle     = "Tools and Algorithms for the Construction and Analysis of Systems",
  year          = "2017",
  publisher     = "Springer",
  pages         = "58--75",
  isbn          = "978-3-662-54577-5"
}

@inproceedings{clarke2000cegar,
  author        = "Clarke, Edmund and Grumberg, Orna and Jha, Somesh and Lu, Yuan and Veith, Helmut",
  title         = "Counterexample-Guided Abstraction Refinement",
  booktitle     = "Computer Aided Verification -- CAV 2000",
  year          = "2000",
  publisher     = "Springer",
  pages         = "154--169",
  isbn          = "978-3-540-45047-4"
}

@article{mccormick1976,
  author  = {McCormick, Garth P.},
  title   = {Computability of global solutions to factorable nonconvex programs: {Part I} --- {Convex} underestimating problems},
  journal = {Mathematical Programming},
  year    = {1976},
  volume  = {10},
  number  = {1},
  pages   = {147--175},
}

@InProceedings{Dutertre:cav2014,
  author =      {Dutertre, Bruno},
  title =      {Yices 2.2},
  booktitle = {Computer-Aided Verification (CAV'2014)},
  year =      2014,
  volume =       8559,
  series =       {LNCS},
  pages =        {737--744},
  publisher =    {Springer}}

@inproceedings{mathsat5,
  author = {Alessandro Cimatti and Alberto Griggio and Bastiaan Schaafsma and Roberto Sebastiani},
  title = {{The MathSAT5 SMT Solver}},
  booktitle = {Proceedings of TACAS},
  year = {2013},
  volume = {7795},
  series = {LNCS},
  publisher = {Springer},
}

@inproceedings{mann2021smt,
  title={{SMT}-switch: a solver-agnostic {C++ API} for {SMT} solving},
  author={Mann, Makai and Wilson, Amalee and Zohar, Yoni and Stuntz, Lindsey and Irfan, Ahmed and others},
  booktitle={Theory and Applications of Satisfiability Testing, {SAT}},
  pages={377--386},
  year={2021},
  organization={Springer}
}
